\documentstyle[prl,aps,twocolumn,epsf]{revtex}
\begin {document}

\twocolumn[\hsize\textwidth\columnwidth\hsize\csname
@twocolumnfalse\endcsname

\large
\parindent 0 cm
\begin {center}
{\bf Internal Avalanches in a Granular Medium} \\
\vskip 0.2 cm
\normalsize
S. S. Manna$^{1*}$ and D. V. Khakhar$^{2\dagger}$\\
$^1$Satyendra Nath Bose National Centre for Basic Sciences,
Block-JD, Sector-III, Salt Lake, Calcutta-700091, India \\
$^2$Department of Chemical Engineering,
Indian Institute of Technology, Powai, Mumbai - 400076, India \\
$^*$manna@boson.bose.res.in,$\quad$$^\dagger$khakhar@cupid.che.iitb.ernet.in\\
\end {center}
\normalsize
\begin {abstract} 
Avalanches of grain displacements can be generated by creating local
voids within the interior of a granular material at rest in a bin.
Modeling such a two-dimensional granular system by a collection of
mono-disperse discs, the system on repeated perturbations, shows all
signatures of Self-Organized Criticality. During the propagation of
avalanches the competition among grains creates arches and in the 
critical state a distribution of arches of different sizes is 
obtained. Using a cellular automata model we demonstrate that the
existence of arches determines the universal behaviour of the model 
system.
\end {abstract}
\centerline{PACS numbers: 05.70.Jk, 64.60 Lx, 74.80.Bj, 46.10.+z}

\vskip2pc]

   The search for self organized criticality (SOC) [1] in granular
systems in general and in sandpiles in particular has been a subject 
of active research over the last decade. How such a system reacts in 
the form of fast cascades of grain displacements, called `avalanches', 
in response to a slow external drive at the microscopic level is the 
crucial question of study. It has been suggested that starting from 
an arbitrary initial condition, a non-equilibrium critical state, 
characterised by scale free avalanches in both space and time, should 
be expected after a long time [1,2].  Other naturally occuring physical 
phenomena like forest fires [3], river networks [4], earthquakes [5] 
etc. have also been proposed as examples of systems showing SOC.

   It was observed that a steady shaped sandpile grown on a fixed base
fulfills all the requirements of SOC. In this state, dropping even a 
few grains creates rapidly moving avalanches of sand sliding along 
the surface of the pile. It was expected that the avalanches should 
be power law distributed in their spatial as well as temporal extents
and therefore a sandpile should be considered as a simple example 
showing SOC [1,2].

   Experimental observations, however, show partial support to this
idea. Sand is allowed to fall from a slowly rotating semi-cylindrical 
drum through the space between the plates of a vertical 
parallel plate capacitor. The Fourier spectrum of the time series data 
of fluctuating capacitance showed a peak contrary to the expectation of 
a power law [6]. Similarly sand dropping from the edge of a sandpile on 
a fixed base was directly measured, and was seen to have a power law 
distribution only for small systems but not for large systems [7]. It 
was argued that due to the approximately spherical shape of the grains 
used in these experiments the effect of inertia cannot be neglected and 
this was the reason for absence of scaling behaviour. This is verified 
in an experiment using rice grains, which are highly anisotropic, and 
criticality was observed [8].

   A number of theoretical models, generally known as `sandpile' models
have been studied. The models are based on stochastically driven cellular 
automata which evolve under a nonlinear, diffusive mechanism leading to 
a nonequilibrium critical state. The most widely studied is the Abelian 
sandpile model (ASM) where the stable configuration does not depend on 
the sequence of sand grain additions [9]. Other variants of the sandpile 
model include situations where the stability of a sand column depends on 
the local slope or the Laplacian of the height profile [10]. A two-state
sandpile model with stochastic evolution rules is also studied [11].

In all the studies discussed above the avalanches propagate on the 
surfaces of the sandpiles. However, there exists the possibility of
creating avalanches in the interior of a granular material. In a 
granular material kept in a bin at rest, different grains support 
one another by mutually acting balanced forces. Now if a grain is 
removed, the grains which were supported by it become unstable and 
tend to move. Eventually the grains in the further neighbourhood 
also loose their stability. As a result an avalanche of grain 
displacements takes place, which stops when no more grains
remain unstable. The basic physical behavior here is thus quite 
different from the avalanches on the surface of the pile because 
of the constraints to particle motion in the dense particle beds.

   A two-dimensional semi-lattice model was studied for this problem 
[12]. Non-overlaping unit square boxes model the grains, whose 
horizontal coordinates can vary continuously where as the vertical 
co-ordinates are discretized. A grain can only fall vertically
if insufficiently supported and sufficient space below is available.
The system is disturbed by removing grains at the bottom and thus
creating avalanches of grain movements.

   During the propagation of avalanches inside a granular medium
grains compete locally with one another to occupy the same vacant space.
The high packing densities of the particle beds prevent a single particle
from occupying the available void space, and consequently particles
get locked to form `arches' [13]. A stable arch is a chain of grains 
where the weight of each grain is balanced by the reaction forces from 
two neighbouring grains in the chain. Arches can form only when a grain
is allowed to roll over other grains. Since the rolling motion of the 
grains was absent therefore the arches were not formed in the granular 
patterns [12].

   We here study a more realistic model of this problem in two dimension, 
where both fall and roll motions of grains are allowed. The granular 
system is represented by $N$ hard monodisperse discs of radii $R$. No 
two grains are allowed to overlap but can touch each other and one can 
roll over the other if possible without slipping. A rectangular area of 
size $L \times L$ on the $x-y$ plane represents our two dimensional bin. 
Periodic boundary condition is used along the $x$ direction and the 
gravity acts along the $-y$ direction. The bottom of the bin at $y=0$ is
highly sticky and any grain which comes in contact gets stuck there
and does not move further.

   The dynamical evolution of the system is studied by a
`pseudo-dynamics' [14]. Unlike the method of molecular dynamics
we donot solve here the classical equations of motion for the
grain system. Only the direction of gravity and the local geometrical
constraints due to the presence of other grains govern the movement
of a grain. For justification of using the pseudo-dynamics
we argue that due to the high compactness of the system
a grain never gets sufficient time to accelerate much. Therefore in our
model, a grain can have only two types of movements in unit time:
The vertical $fall$ up to a maximum distance $\delta$ and
the $roll$ up to a maximum angle $\theta=\delta/2R$ over another
grain in contact.

   Movement of a grain needs the information on the positions of other
grains in the neighbourhood which it may possibly interact. An efficient
way to search this is to digitize the bin into a square grid and associate
the serial number of a grain to the primitive cell of the grid
containing its centre. Choise of $R=1/\sqrt 2 +$ ensures
that a cell corresponds to at most one grain. With this choice
it is sufficient to search only within the 24 neighbouring cells for
possible contact grains. The weight of a grain $n$ is supported by
two other grains. If $n_L$ and $n_R$ are the serial numbers of the left
$(L)$ and right $(R)$ supporting grains, then the grain $n$ is updated as:
(i) If $n_L=n_R=0$, it falls,
(ii) if $n_L \ne 0$ but $n_R = 0$, it rolls over $n_L$,
(iii) if $n_L = 0$ but $n_R \ne 0$, it rolls over $n_R$,
(iv) if $n_L \ne 0$ and $n_R \ne 0$ it is stable.

   When the grain $n$ with the centre at $(x_n,y_n)$ is allowed to fall,
it may come in contact with a grain $r$ at $(x_r,y_r)$ within the distance
$\delta$. The new coordinates are:
\[
y_n'=y_r+\sqrt{4R^2-(x_n-x_r)^2} \quad {\rm if} \quad y_n-y'_n < \delta \quad 
\]
\vskip -1.2cm
\[
{\rm otherwise,} \quad y'_n=y_n-\delta.
\]
Similarly the grain $n$, while rolling over the grain $r$, may
come in contact with another grain $t$ at $(x_t,y_t)$. The new coordinates
for the centre of $n$ where it touches both $r$ and $t$ are:
\[
x_n' = \frac{1}{2}(x_t+x_r) +
g(y_t-y_r)\sqrt{\frac{4R^2}{d_{rt}^2}-\frac{1}{4}}
\]
\[
y_n' = \frac{1}{2}(y_t+y_r) -
g(x_t-x_r)\sqrt{\frac{4R^2}{d_{rt}^2}-\frac{1}{4}}.
\]
Here $d_{rt}$ is the distance between the centres of the grains $r$ and $t$
and the $g=+1$ and -1 for left and right rolls.
To reach the new position if the grain $n$ rolls an angle $\theta_m < \theta$
then it is accepted, otherwise it rolls up to an angle $\theta$.

   The initial grain pattern is generated by the ballistic deposition method
with restructuring [15,16].  Grains are released sequentially one after
another along randomly positioned vertical trajectories and are allowed
to fall till they touch the growing pile and then roll down along the
paths of steepest descent to their local stable positions. We notice
that in the initial pattern no big arch exists since a grain while rolling down
along the surface does not need to compete with any other grain.
Using a system of $L$=80, we compute the packing fraction $\rho=$ 0.822 $\pm$
0.005
consistent with more precise estimate of 0.8180 $\pm$ 0.0002 in [16].
\vskip -4.6cm
\begin{figure}[h]
\centerline{\epsfxsize=300pt\epsfbox{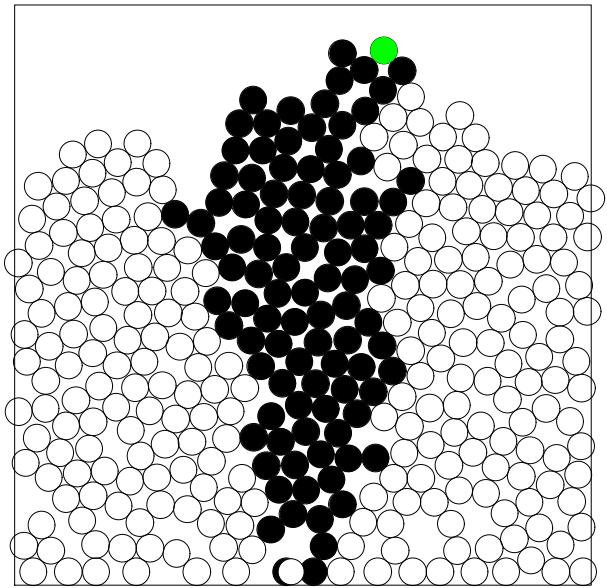}}
\vskip -2.0cm
\caption
{
Picture of a typical avalanche. The vacant circles
denote the undisturbed grains, filled circles denote the displaced grains,
the opaque circle at the bottom denotes the position of the grain
which was removed and the shaded circle denote the position of the
grain where the removed grain was replaced. In a system of size
$L$=30 and number of grains $N$=340, 97 grains took part in the
avalanche.
}
\end{figure}

   The system is repeatedly perturbed by removing grains randomly
at the bottom one after the other. Every time a grain is removed,
an avalanche is followed and after the avalanche is over the removed
grain is placed back at a random position on the top surface
again using the same ballistic method (Fig. 1).
We first observe that $\rho$ of the stable configuration, averaged over many
initial random patterns generated with the same algorithm, decreases
with the number of avalanches created and finally reaches a steady
value of 0.748 $\pm$ 0.005. A similar study but with different initial
configurations with a different value of average initial $\rho$,
shows the final steady state packing fraction reaches the same value,
which implies that the final stable state is independent of the initial state.
The total number of grain displacements is called the avalanche
size $s$ and the duration of the avalanche is the life time $t$.
Power law distributions are observed for $s$ and $t$:
$D(s) \sim s^{-\tau_s}$ and $D(t) \sim t^{-\tau_t}$
with $\tau_s \approx 1.7$ (Fig. 2) and $\tau_t \approx 2.1$.
\vskip -4.0cm
\begin{figure}[h]
\centerline{\epsfxsize=250pt\epsfbox{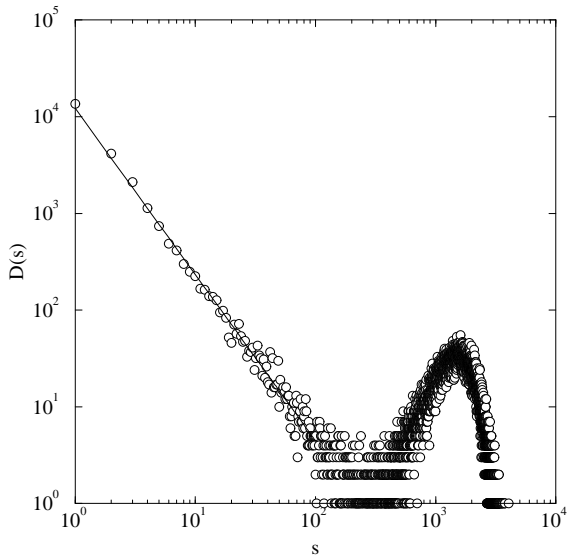}}
\vskip -1.0cm
\caption
{
Plot of the avalanche size distribution $D(s)$ vs. $s$
for a system of size $N$ = 10000 and over 70000 avalanches. The
straight part fits with a slope $\tau_s$ = 1.7.
}
\end{figure}

   The observed exponents in this model are significantly larger than
the corresponding values of $\tau_s \approx 1.34$ and $\tau_L \approx 1.47$ in
[12].
We argue that the reason for this difference could be the existence of
arches in our model. Due to the arches, the propagation of avalanches
get arrested more frequently than in [12] and therefore smaller avalanches are
more probable which enhances the values of the critical exponents in our model.

   To demonstrate more explicitly that the above reasoning may be
correct, we study a cellular automata
model for the granular systems. A square lattice of size $L$ with periodic
boundary condition along the $x$ axis represents the bin. The gravity acts
in the $-y$ direction. Positions of the grains are limited only to the lattice
sites: a site can be either occupied by a grain or remain vacant.
Initially the bin is filled up again using the ballistic method.

   A single movement of a grain at $C(i,j)$ involves the neighbouring seven
sites: $LU(i-1,j+1), L(i-1,j), LD(i-1,j-1), D(i,j-1), RD(i+1,j-1), R(i+1,j)$
and
$RU(i+1,j+1)$.In the $fall$ move the grain comes down one level to the vacant
site at
$D$ and in $roll$ move the grain goes either to the vacant site at $LD$ or at
$RD$
(Fig. 3). An arch
is formed when a grain at $C$ is considered stable if both of its two
diagonally
opposite sites either at $LD$ and $RU$ or, at $LU$ and $RD$ are occupied.
As a result, on the square lattice, the only possible shape of an arch
consists of two sides of a triangle. Depending on whether we allow the
arch formations or not, we define the following two models.
\vskip -5.0cm
\begin{figure}[h]
\centerline{\epsfxsize=300pt\epsfbox{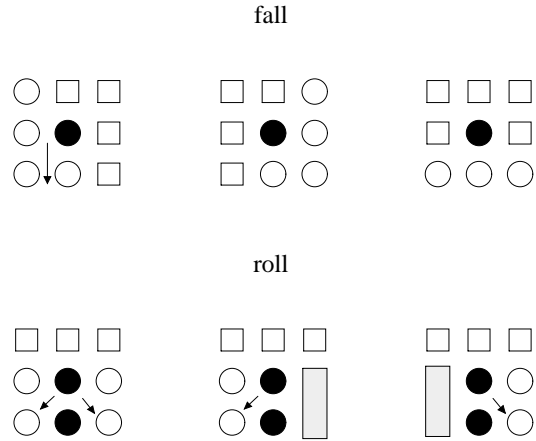}}
\vskip -2.7cm
\caption
{
The possible $fall$ and $roll$ moves in the cellular
automata model of the granular system on the square lattice.
Filled circle denotes the position of a grain, unfilled circle
denotes a vacant site. The grain moves to the vacant position
irrespective of the occupation of the sites with square boxes.
Shaded rectangle denotes a pair of sites, in which at least
one is occupied.
}
\end{figure}

   In model A we allow arch formations. The grain at $C$ falls only if any of
the
following three conditions is satisfied:
(i)   $LD$, $L$ and $LU$ are vacant
(ii)  $RD$, $R$ and $RU$ are vacant
(iii) $LD$ and $RD$ are vacant.
In all other situations the grain does not fall. Notice that in conditions (i)
and (ii) we are allowing the formation of arches.
The grain at $C$ rolls only if the site $D$ is occupied. This is done in any of
the
three following ways:
(i) If both $LD,L$ are vacant but either of $RD,R$ is occupied then the grain
rolls to $LD$.
(ii) if both $RD,R$ are vacant but either of $LD,L$ is occupied then the grain
rolls to $RD$.
(iii) If all four sites at $LD, L, RD, R$ are vacant then the grain rolls
either to $LD$ or to $RD$ with
probability $1/2$.
A steady state pattern is shown in Fig. 4.

   In model B we do not allow arch formations. The first two conditions for
fall of model $A$ are modified as:
(i)  $LD$ and $L$ are vacant
(ii) $RD$ and $R$ are vacant. All other conditions of fall as well as roll
remain same as in model A.

   Initial granular patterns are generated using the same random ballistic 
deposition method with restructuring [15,16] and patterns are same for both 
models.  As before the avalanches are created by taking out one grain at a 
time at the bottom, allowing the system to relax and replacing it randomly 
on the surface after the avalanche is over. Here also we see that the average 
density of sites, starting from
an initial value of $0.907 \pm 0.005$, decreases to the final stable
value of $0.590 \pm 0.005$ in model A and to $0.618 \pm 0.005$ in model B.
The avalnche size $s$ and life time $t$ follow power law distributions:
$D(s) \sim s^{-\tau^A_s}, D(t) \sim t^{-\tau^A_t}$ and similarly for the 
model B. Different exponents are obtained for the two models:
$\tau^A_s \approx 1.48$  and $\tau^A_t \approx 1.99$ where as
$\tau^B_s \approx 1.34$ and $\tau^B_t \approx 1.50$. We explain that
absence of arches makes the exponent values for the model B close
to that of [12] but their presence enhances the values in model A.
\vskip -5.0cm
\begin{figure}[h]
\centerline{\epsfxsize=330pt\epsfbox{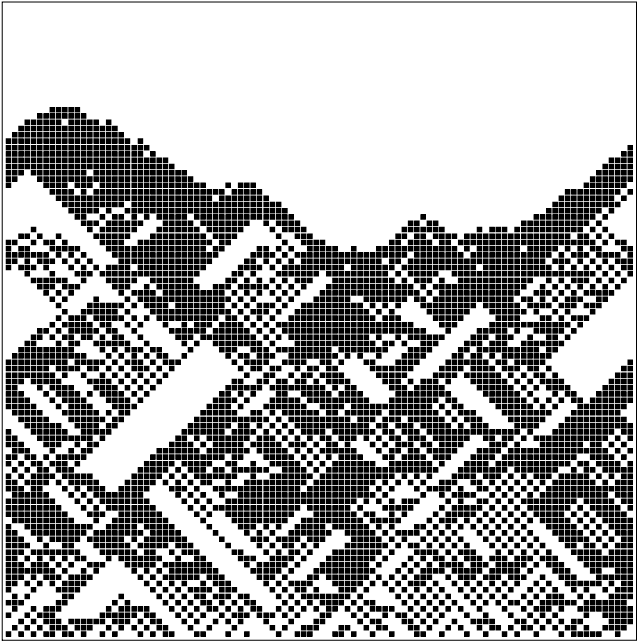}}
\vskip -1.9cm
\caption
{
Figure 4: A stable configuration of the granular system in the
cellular automata model A after a large number of avalanches
are created. Triangular arches are noticed.
}
\end{figure}

To summarise, avalanches of grain displacements can be created
in the interior of a granular material at rest by locally disturbing
the system. In a numerical study we provide indications that
on repeated creations of such avalanches, the granular system
reaches a critical state characterised by long range
correlations. The presence of arches play an important role
in determining the critical behaviour. Since in an avalanche
grain motions are highly constrained, the effect of inertia may not be
very significant. Therefore, we conjecture that in real experiment
of internal avalnches it should be possible to observe SOC even with spherical
grains unlike the case of surface avalanches,
where anisotropic grains were necessary to observe criticality.

The funding provided by the Indo-French Centre for the Promotion of
Advanced Research (IFCPAR) is gratefully acknowledged.


\parindent= 0 cm
\vskip 0.2 cm

\begin {itemize}

\item[[1]] P. Bak, C. Tang and K. Wiesenfeld, Phys. Rev. Lett. {\bf 59}.
          381 (1987).

\item[[2]] P. Bak, C. Tang and K. Wiesenfeld, Phys. Rev. A {\bf 38}, 364
          (1988); P. Bak, {\it How Nature Works: The Science of
          Self-Organized Criticality}, (Copernicus, New York, 1996).

\item[[3]] P. Bak and K. Chen, Physica D {\bf 38}, 5 (1989),
           H-M Br\"oker and P. Grassberger, Phys. Rev. E {\bf 56},
           R4918 (1997).

\item[[4]] A. Rinaldo, I. Rodriguez-Iturbe, R. Rigon, E. Ijjasz-Vasquez
          and R. L. Bras, Phys. Rev. Lett. {\bf 70}, 822 (1993);
          S. S. Manna and B. Subramanian, Phys. Rev. Lett. {\bf 76}
          (1996) 3460.

\item[[5]] J. M. Carlson and J. S. Langer, Phys. Rev. Lett. {\bf 62},
          2632 (1989); Z. Olami, H. J. S. Feder and K. Christensen,
          Phys. Rev. Lett. {\bf 68}, 1244 (1992).

\item[[6]] H. M. Jaeger, C-h Liu and S. R. Nagel, Phys. Rev. Lett.
           {\bf 62}, 40 (1989).

\item[[7]] G. A. Held, D. H. Solina II, D. T. Keane, W. J. Haag, P. M. Horn
          and G. Grinstein, Phys. Rev. Lett. {\bf 65}, 1120 (1990).

\item[[8]] V. Frette, K. Christensen, A. Malthe-Sorenssen, J. Feder,
          T. Jossang and P. Meakin, Nature (London) {\bf 379}, 49 (1996).

\item[[9]] D. Dhar, Phys. Rev. Lett. {\bf 64}, 1613 (1990).

\item[[10]] L. P. Kadanoff, S. R. Nagel, L. Wu and S. Zhou, Phys. Rev. A.
          {\bf 39}, 6524 (1989); S. S. Manna, Physica A {\bf 179}, 249
          (1991).

\item[[11]] S. S. Manna, J. Phys. A {\bf 24}, L363 (1992).

\item[[12]] R. E. Snyder and R. C. Ball, Phys. Rev. E {\bf 49}, 104
            (1994).

\item[[13]] D. E. Wolf, in {\it Computational Physics} ed. by K. H.
Hoffmann and M. Schreiber, Springer 1996.

\item[[14]] S. S. Manna and D. V. Khakhar, in
      {\it Nonlinear Phenomena in Material Science III},
      G. Ananthakrishna, L. P. Kubin and
      G. Martin (Eds.) (Transtech, Switzerland, 1995).

\item[[15]] W. M. Visser and M. Bolsterli, Nature {\bf 239}, 504 (1972).

\item[[16]] R. Jullien, P. Meakin and A. Pavlovitch, in
            {\it Disorder and Granular Media}, D. Bideu and A. Hansen
            (Eds.) (Elsevier 1993).

\end {itemize}

\end {document}